# Giant enhancement in Rashba spin-Seebeck effect in NiFe/p-Si thin films

Ravindra G. Bhardwaj[1], Paul C Lou[1] and Sandeep Kumar[1,2*]

[1] Department of Mechanical Engineering, University of California, Riverside, CA 92521, USA

[2] Materials Science and Engineering Program, University of California, Riverside, CA 92521, USA

Abstract

The spin-Seebeck effect mediated thermoelectric energy conversion can provide efficient alternative to traditional thermoelectrics for waste heat recovery. To achieve this goal, efficient spin to charge conversion using earth-abundant materials is essential. Proximity induced Rashba effect arises from the charge potential mediated by structural inversion asymmetry, which has been reported in Si thin films and can be manipulated by controlling the thickness of Rashba layer. We demonstrate a giant Rashba spin-Seebeck effect in NiFe/p-Si (polycrystalline) bilayer thin films. The bilayer thin film specimens have p-Si layer thickness of 5 nm, 25 nm and 100 nm while keeping the NiFe layer thickness at 25 nm. The Rashba spin Seebeck coefficient has been estimated to be $0.266 \frac{\mu V}{K}$ for 100 nm p-Si, and increases by an order of magnitude to $2.11 \frac{\mu V}{K}$ for 5 nm p-Si. The measured spin-Seebeck coefficient in 5 nm p-Si specimen is one of the largest coefficient ever reported. The measured voltage of 100.3 µV is one of the largest reported spin-Seebeck voltage, with smallest area of ~160 X 10 µm² used in any spin-Seebeck measurement. This scientific and technological breakthrough using earth abundant elements brings the spin mediated thermoelectric energy conversion for waste heat recovery closer to reality.

Since the discovery of spin-Seebeck effect (SSE) by Uchida et al.[1], the spin mediated thermoelectric energy conversion has been extensively investigated for ferromagnetic metals, ferromagnetic insulators, antiferromagnetic materials and oxides[2-9]. The SSE is an interface effect that generally occurs between a spin polarized ferromagnet(FM) and a normal metal(NM). In SSE, the thermal transport takes place due to the two-step spin dependent process. In the first step, the thermal gradient leads to generation of heat current from the phonon-magnon or phonon-electron interactions[10-11]. The heat current leads to generation of spin currents in the spin-polarized material[12]. Spin current is in the form of either magnons or spin-polarized current due to electron or combination of both[11]. In the second step, spin current is injected across the interface from FM into NM due the spin potential gradient between FM and NM at the interface. The spin to charge conversion takes place in the NM (usually heavy metal) due to inverse spin-Hall effect. The FM is spin source and the NM is the spin sink[3]. Thus, the advantage of SSE compared to conventional thermoelectric effect is that it uses the properties of two or more materials that can be independently optimized[13]. The spin Seebeck effect and inverse spin-Hall effect (ISHE) produce an electric field given by,

$$E_{ISHE} = -S\sigma \times \nabla T \qquad (1)$$

where, S is spin Seebeck coefficient and $\sigma$ is the spin polarization vector. Since the above equation is similar to equation of anomalous Nernst effect (ANE) voltage (where $\sigma$ is replaced by *M* (magnetization)) and thus both SSE and ANE have identical symmetry. One of the controlling factors in spin mediated thermoelectric energy conversion is spin to charge conversion due to spin-orbit coupling in NM. The spin Hall Angle (SHA) is the measure of efficiency in conversion of charge current to spin current given by ratio of generated charge current to the injected spin current[14] and vice versa. Pt is the primary material for spin to charge conversion due to its large SHA, which can be enhanced by defects and impurities. Extensive research has been reported in methods to enhance the spin-Hall angle for inverse spin Hall effect. These methods include alloying[15] and metastable phases[16-17].

Recently, Bhardwaj et. al.[18] reported SSE and thermal spin galvanic effect (SGE) in $Ni_{80}Fe_{20}$ (poly) bilayer specimen without any heavy metal detector. They propose that the spin to charge conversion in p-Si layer in the bilayer specimen is due to structure inversion asymmetry of sandwich structure and proximity effect. They observed that the spin-Seebeck coefficient in the bilayer is of the same order as Pt. However, larger values of spin Seebeck Coefficient ($S_{LSSE}$) is required to make the efficient spin mediated thermoelectric technologies into reality. The Rashba spin-orbit coupling (SOC) relies on the charge potential due to structure inversion asymmetry (SIA), which can be controlled by reducing the thickness of the sandwiched layer. The Rashba effect mediated spin-Hall magnetoresistance has also been reported in $Ni_{81}Fe_{19}$/MgO/p-Si[19-21] and $Ni_{81}Fe_{19}$/MgO/n-Si[22] thin films as well. The spin-Hall magnetoresistance arises due to ISHE, which is essential for SSE. We hypothesize the reduction in thickness of p-Si will increase Rashba SOC, leading to efficient spin to charge conversion. To explore the thickness dependent SSE behavior, we fabricated three specimen having p-Si layer thickness of 5nm, 25nm and 100nm while keeping the thickness of NiFe at 25 nm. In this work, we demonstrate giant enhancement in spin mediated thermoelectric energy conversion due to efficient spin to charge conversion from Rashba SOC.

The SSE is usually characterized using two universal device configurations- longitudinal spin Seebeck effect (LSSE) and transverse spin Seebeck effect (TSSE)[23]. Spin current is parallel to the temperature gradient in LSSE[24] while it is perpendicular to the temperature gradient in TSSE[9, 25-26]. In this work, we use the LSSE configuration to discover the spin mediated thermoelectric energy conversion behavior in NiFe/p-Si bilayers as shown in Figure 1 (a). In the LSSE configuration, the temperature gradient across the thin film specimen creates a spin current ($J_s$), which then get converted into a charge current ($J_c$) as shown in Figure 1 (a). To fabricate the experimental setup, we take a Si wafer and deposit 450 nm of thermal silicon oxide using chemical vapor deposition (CVD). We, then, deposit the NiFe/p-Si (poly) bilayers using the RF sputtering as shown in Figure 1 (b). We deposit three sets of bilayers with 5 nm, 25 nm and 100 nm of p-Si while keeping the NiFe thickness at 25 nm. We hide three quarters of the wafer and deposit each of the bilayer individually. The p-Si target is B-doped having resistivity of 0.005-0.01 Ω-cm. We sputter 50 nm MgO to electrically isolate the heater and the specimen. We then deposit heater composed of Ti (10 nm)/Pt (100 nm). The insulator and heater deposition are common to all the bilayers devices, which reduces the fabrication induced measurement variations. We acquire experimental data inside a Quantum design Physical property measurement system (PPMS). To ascertain the thermal response characteristics, we acquire the second harmonic response as a function of current at an applied magnetic field of 1500 Oe as shown in Figure 1 (c). We observe that the $V_{2\omega}$ response shows a relationship having both quadratic and linear terms with respect to the applied heating current. This behavior suggests that the temperature rise across the specimen do not have linear relationship with the square of heating current, which may be due to temperature drop across MgO (top) and thermal oxide layers (bottom).

(Figure 1)

To investigate the SSE behavior, we acquire the $V_{2\omega}$ response as a function of magnetic field (1500 Oe to -1500 Oe) applied in the y-direction (normal to the temperature gradient) for all the three specimens. The data is acquired at 10 K, 100 K and 300 K as shown in Figure 2 a-c.

The observed behavior demonstrates the magnetic switching behavior for all the thicknesses, which is consistent with SSE and ANE. At 300 K, the $V_{2\omega}$ for the 5 nm, 25 nm and 100 nm are 100.3 µV, 36.77 µV and 31.08 µV respectively. Although the $V_{2\omega}$ response for the 25 nm and 100 nm specimens is similar in magnitude but the magnitude of $V_{2\omega}$ response for 5 nm is extremely large. This behavior eliminates ANE as an underlying mechanism for the observed behavior since the NiFe layer thickness is same across all the specimens. For specimen with 5 nm p-Si layer thickness, the $V_{SSE}$ is 100.3 µV at 300 K, which is significantly larger as compared to any other SSE measurement reported in the literature. In addition, the SSE specimen area is 160X10 µm$^2$ in this study, which is an order of magnitude smaller than the other reported experiments. Notable, this efficient spin mediated thermoelectric energy conversion is achieved without using any heavy metal for spin to charge conversion. This giant enhancement in SSE, presented in this study, is attributed to the proximity induced Rashba SOC in p-Si layer, which increases significantly with reduction in p-Si layer due to structure inversion asymmetry, resulting in the observed behavior. The Rashba SOC may also give rise to spin-galvanic effect (SGE). In the recent work, Bhardwaj et al.[18] reported thermal SGE for an out of plane magnetic field, where its magnitude is reported to be similar to SSE. Here, we undertake a similar experiment to uncover the thermal SGE behavior. We measure the $V_{2\omega}$ response as a function of magnetic field (from -2500 Oe to 2500 Oe) applied in the z-direction (parallel to temperature gradient) at 300K for a heating current a 20mA as shown in the Figure 2 d. The observed $V_{2\omega}$ response is similar to SSE measurement. The $V_{2\omega}$ response in case of specimen having 5 nm p-Si layer is very large as compared to specimens having 25nm and 100nm p-Si thick layer. We propose that this out of plane $V_{2\omega}$ response is due to thermal SGE.

(Figure 2)

The measurement of $V_{2\omega}$ response as a function of magnetic field leads to confirmation of the SSE and thermal SGE behavior. To uncover the underlying mechanism of SSE mediated energy conversion behavior, we measure $V_{2\omega}$ response as a function of temperature from 350 K to 10 K under a magnetic field of 1000 Oe applied along the y-direction. We observe a gradual decrease in the $V_{2\omega}$ response as the temperature is lowered to 10 K as shown in Figure 3 a. The temperature dependent behavior signifies that the observed $V_{2\omega}$ response is due to magnon mediated SSE. To further support our argument, we measured angular dependence of the $V_{2\omega}$ response for a constant applied magnetic field of 2 T rotated in the yx-plane. For all the specimens, we observe an embeded cosine behavior attributed to SSE. The deviation from the cosine behavior can arise from the thermal SGE contributions.

(Figure 3)

To quantify the SSE in this study, we estimate the longitudinal spin Seebeck coefficient as[18, 27]

$$S_{LSSE} = \frac{E_{ISHE}}{\nabla T} = \frac{V_{ISHE} t_{FM}}{w_{NM} \Delta T} \qquad (2)$$

where, $\Delta T$ is the temperature gradient across the sample, $w_{NM}$ is the distance between electrical contact in Normal Metal (NM) (~160 $\mu m$), $t_{FM}$ is the ferromagnet (FM) material thickness, and $V_{ISHE}$ is the $V_{2\omega}$ response measured due to ISHE in NM. Using 3ω method, we can experimentally calculate the temperature gradient between heater and the far field substrate temperature. The temperature gradient between heater and substrate using 3ω method[28] is given by

$$\Delta T = \frac{4V_{3\omega}}{R' I_{rms}} \tag{3}$$

Where, $V_{3\omega}$ is the measured third harmonic response, $R'$ is the slope of the resistance as a function of temperature and $I_{rms}$ is the applied heating current. The heater temperature is estimated to be ~313.7 K for bilayer specimen having 25 nm p-Si layer and 313.4 K in case of 100 nm p-Si specimen for $R'$ of 0.07 Ω/K[18]. We use finite element method (FEM) (COMSOL software) to simulate the temperature gradient across the bilayer specimen, which is essential for spin-Seebeck coefficient. For modeling the temperature gradient, we assumed the $\kappa_{p-Si}$ to be 15 W/mK, 20 W/mK and 35 W/mK [29-30] for thickness of 5 nm, 25 nm and 100 nm respectively and $\kappa_{NiFe}$ to be 20 W/mK [31]. We observe that simulated temperature gradient across the NiFe layer is similar for all the p-Si layers. This observation reinforces that the ANE is not the underlying reason of observed $V_{2\omega}$ response. Using the simulated temperature gradient across the bilayer, the $S_{LSSE}$ is calculated to be 2.11 $\frac{\mu V}{K}$, 0.506 $\frac{\mu V}{K}$ and 0.266 $\frac{\mu V}{K}$ for 5nm, 25nm and 100nm thick p-Si layer thickness respectively as shown in Figure 3 (c). As stated earlier, the observed SSE behavior is proposed to occur due to proximity Rashba SOC. The Figure 3 (d) shows the mechanism of the observed Rashba SSE behavior. Using ARPES measurement, Gierz et al.[32] demonstrated giant spin-splitting at the Bi/Si (111) interface, which is ascribed to structural inversion asymmetry or Rashba effect. This Rashba energy of 140 meV is reported, which is larger than any other semiconductor heterostructure. In addition, anisotropic electronic band structure is reported for Si (110) using ARPES and simulations[33]. These measurements support our hypothesis. This study includes a ferromagnet having large spin-orbit coupling [34-35], which will introduce in-plane Rashba spin spitting in addition to lifting the degeneracy of band structure due to ferromagnetic proximity effect as show in Figure 3 (d). While Rashba SOC can give rise to giant spin to charge conversion, it may not give rise to ISHE essential for SSE observe in this study. But, Rashba SOC can lift the degeneracy of electronic band structure. The splitting of band structure due to Rashba SOC and ferromagnetic proximity effect can give rise to ISHE. We propose that a combined effect Rashba SOC and ferromagnetic proximity effect leads to the observed giant SSE response in this study. Since, we observe SSE in p-Si layer having thickness of 100 nm, the resulting behavior cannot arise from the two-dimensional electron gas only. The proposed Rashba SOC is bulk, which is consistent with recent reports on Si[19, 22, 36]. With the reduction in p-Si layer thickness, the Rashba SOC will increase and in turn SSE response, which is supported by our measurements.

The thickness dependent LSSE measurement can be used to calculate the spin-Hall angle and spin diffusion length. Qu et al.[37] used the LSSE measurement to uncover the spin-Hall angles using the following equation-

$$\Delta V_{th} \text{ or } V_{SSE} = [2CL\Delta T][\rho(t)\theta_{SH}][\frac{\lambda_{SF}}{t}\tanh\left(\frac{t}{2\lambda_{SF}}\right)] \qquad (4)$$

Where $\Delta V_{th}$ is the thermal voltage due to SSE, the first factor on the right-hand side $[CL\Delta T]$ relates to spin injection efficiency, the length of the wire, and temperature gradient respectively. The second factor on the right-hand side $[\rho(t)\theta_{SH}]$ material specific quantity and relates to the spin conductivity. The last factor relates to the spin diffusion length ($\lambda_{SF}$) and thickness (*t*). Note that the equation 4 assumes that the intrinsic spin diffusion length and spin-Hall angle are independent of material thickness. Although this assumption may be true for the intrinsic spin orbit coupling in case of 5d heavy metals[15, 37] but the Rashba SOC that is responsible for the ISHE is thickness dependent. Hence, the equation 4 cannot quantify the spin transport behavior in NiFe/p-Si bilayer thin films presented in this work.

Instead of quantitative analysis, we undertake a comparative study of the observed SSE in NiFe/p-Si bilayer specimen. We analyzed the reported LSSE measurement over the years for various materials. To demonstrate the quantum of spin-Seebeck behavior observed in the present study, we list the reported $V_{SSE} > 10$ µV in Table 1, where three articles report spin-Seebeck voltage of more than 100 µV[38] [39]. Jiang et al. reported a spin-Seebeck voltage of 175 µV in Bi doped topological insulator ($Sb_2Te_3$), which is probably the highest $V_{SSE}$ reported in the literature. While the $V_{SSE}$ reported in present study is smaller, but the area of the specimen in present study is ~56.25 times smaller as well. In addition, the heating power used to achieve the 175 µV is 500 mW whereas the heating power of 17.6 mW is used in this study to generate 100.3 µV. This observation clearly demonstrates the superiority of the thermoelectric efficiency in the NiFe/p-Si bilayer system. Lin et al. [40] demonstrated a similar $V_{SSE}$ with NiO spacer in between YIG and Pt. They also report the highest SSC of 6 µV/K. The SSC is proportional to the thickness of the FM layer (YIG), which is 500 µm in this reported study. In addition, the specimen area is three orders of magnitude larger as compared to present study. Ramos et al. demonstrated a giant spin-Seebeck voltage in $Fe_3O_4$/Pt system using a spin-Hall thermopile setup. But the specimen area is 3 orders of magnitude larger than the specimen area in this study. In addition, the spin-Hall thermopile configuration can be applied to NiFe/p-Si bilayer system as well to achieve even higher voltages. We have listed other reports of large spin-Seebeck voltage. All specimens in the listed studies consist of areas that are three orders of magnitude larger, with spin-Seebeck voltages that are an order of magnitude smaller than presented in this work.

Table 1. The summary of largest spin-Seebeck voltages and corresponding spin source, spin detector, specimen dimensions and spin-Seebeck coefficient.

| ΔV(μV) | Spin source | Detector | Specimen (L×B) | SSC (μV/K) | Dimensional Normalization | Ref. |
|---|---|---|---|---|---|---|
| ~175 | YIG | $(Bi_xSb_{1-x})_2Te_3$ | 900 μm×100 μm | Not reported | N/A | [38] |
| ~175 | YIG | Pt/NiO | 7 mm×2 mm | 6 | Yes | [40] |
| 100.3 | NiFe | p-Si (poly) | 160 μm×10 μm (smallest) | 0.2-2.2 | Yes | This study |
| ~100 | $Fe_3O_4$ | Pt | Spin-Hall thermopile (7 mm× 2 mm) | Not reported | N/A | [39] |
| ~26 | YIG | Pt | 6 mm×2 mm | 0.100 | No | [26] |
| ~25 and ~12 | $Fe_3O_4$ | Pt | 7 mm×2 mm | 0.03 and 0.7 | Yes | [13, 41] |
| ~18 | YIG | Pt | 10 mm×2.3 mm | 1.500 | No | [42] |
| ~12 | $NiFe_2O_4$ | Pt | 8 mm×5 mm | 0.030 − 0.020 | No | [43] |

In conclusion, we report a giant increase in SSE in NiFe (25 nm)/p-Si (polycrystalline) bilayer specimens having p-Si thickness of 5 nm, 25 nm and 100 nm. The spin-Seebeck voltage shows a three-fold increase in case of 5 nm p-Si specimen as compared to the 25 nm and 100 nm p-Si specimens. The inverse spin-Hall effect is proposed to occur due to proximity induced Rashba spin orbit coupling at the NiFe/p-Si interface. This observation eliminates the requirement of heavy metal (Pt or Ta) for spin to charge conversion. The largest spin-Seebeck coefficient reported in this study is a technological breakthrough, which may help in realization of waste heat recovery applications using spin-Seebeck effect.

**Acknowledgement**

RGB and PCL has equal contribution to this work.

List of Figures-

Figure 1. (a) the schematic showing the longitudinal spin-Seebeck effect measurement setup, (b) the scanning electron micrograph showing the experimental setup, and (c) the second harmonic response as a function of heating current for an applied magnetic field of 1500 Oe.

Figure 2. The longitudinal spin-Seebeck effect measurement for magnetic field applied along the y-direction at 300 K, 100 K and 10 K for the specimen having 25 nm of NiFe layer and p-Si layer thickness of (a) 100 nm, (b) 25 nm and (c) 5 nm, and (d) the thermal spin galvanic effect measurement at 300 K for field applied along z-direction for specimen having p-Si layer thickness of 100 nm, 25 nm and 5 nm of p-Si.

Figure 3. The spin-Seebeck voltage response for bilayer specimen having p-Si layer thickness of 100 nm, 25 nm and 5 nm (a) as a function of temperature from 350 K to 5 K, (b) angular dependence in the yx-plane for an applied magnetic field of 2 T, (c) the calculated spin-Seebeck coefficient as a function of thickness, and (d) schematic of the proposed mechanism for the observed behavior showing proximity Rashba spin splitting and resulting spin to charge conversion.

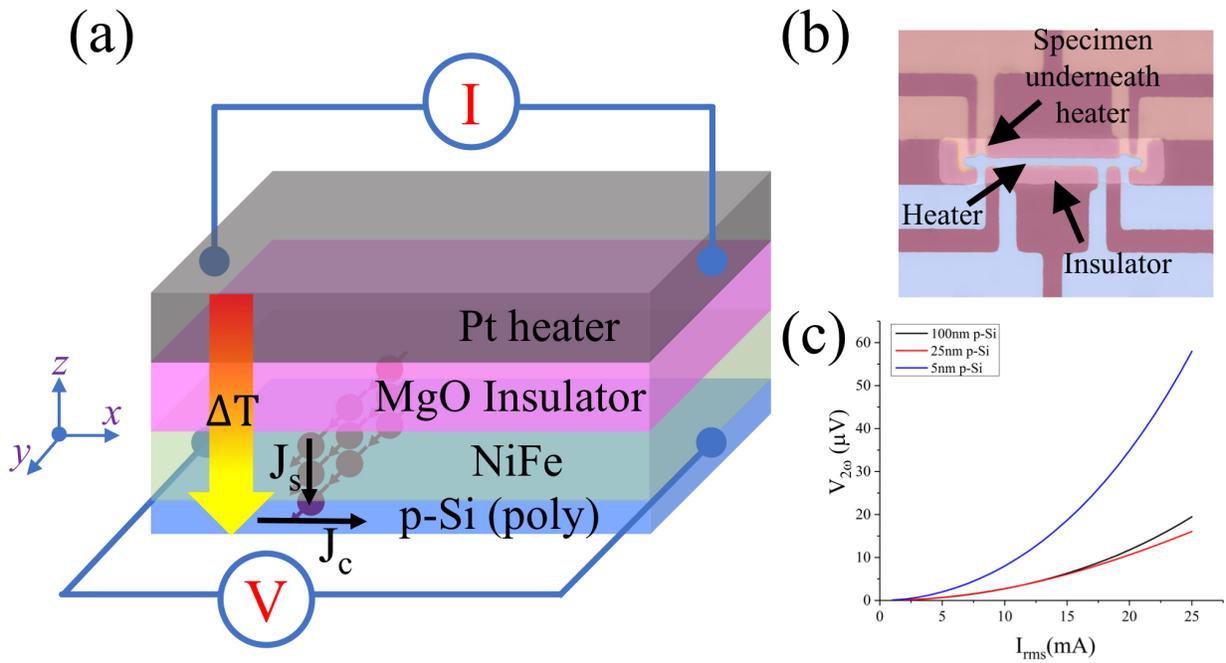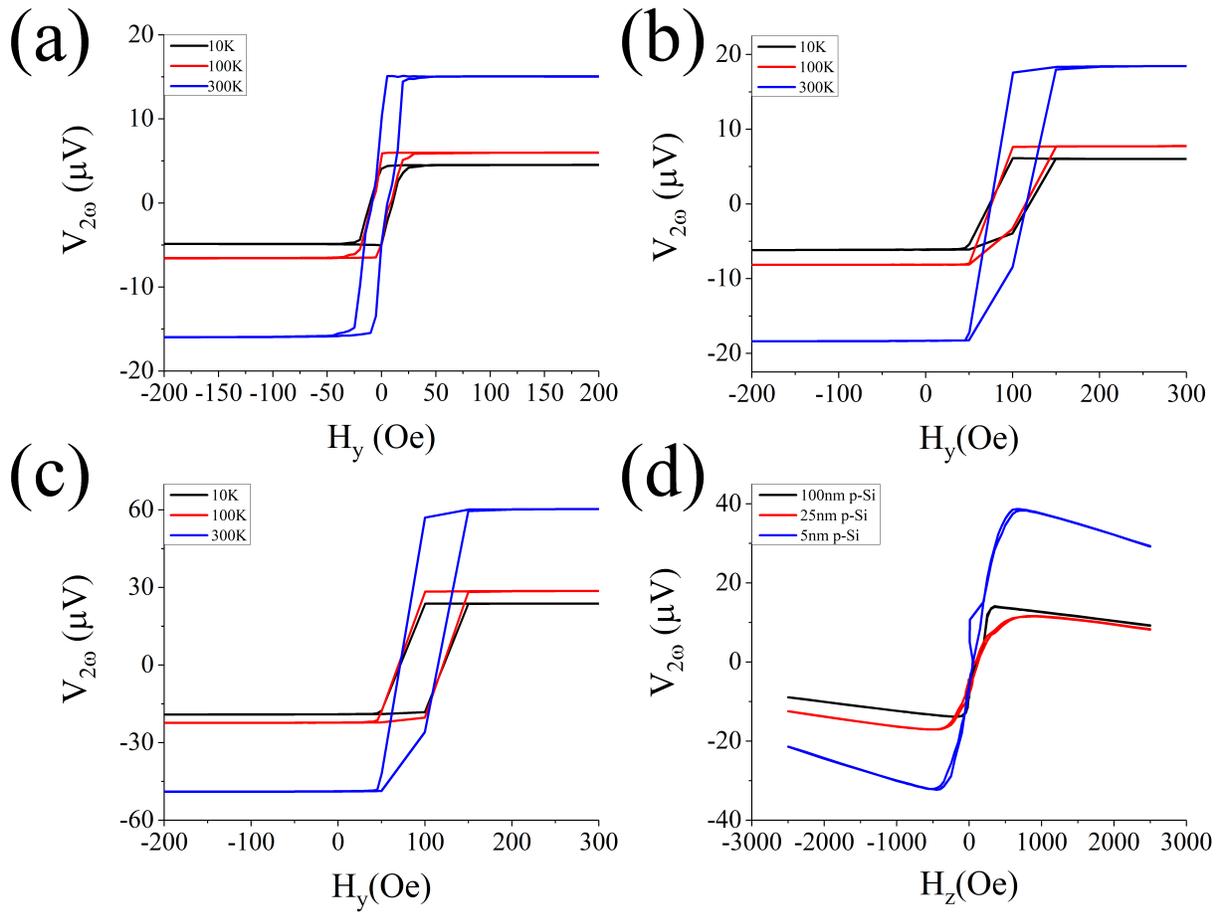

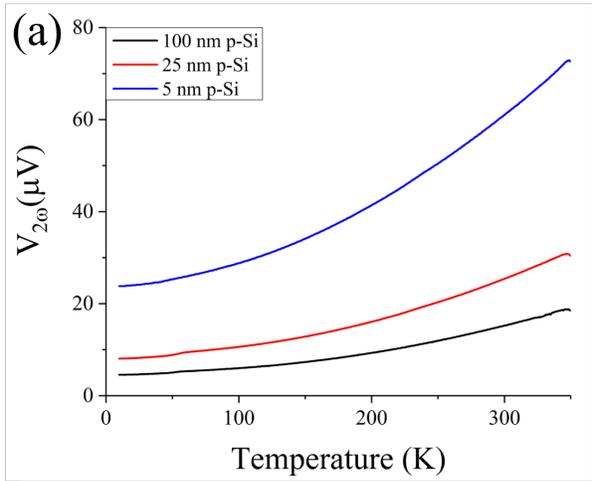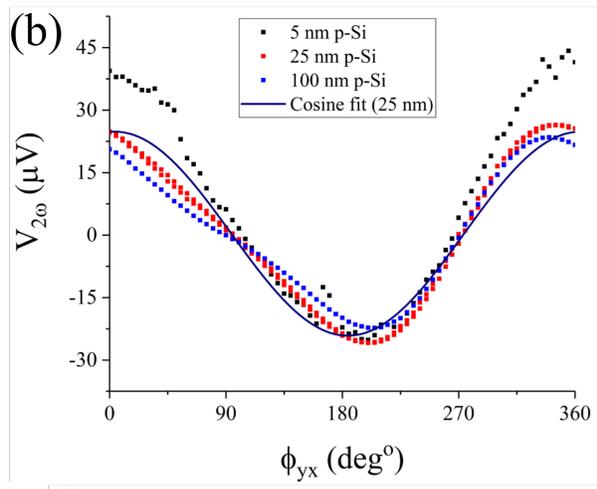
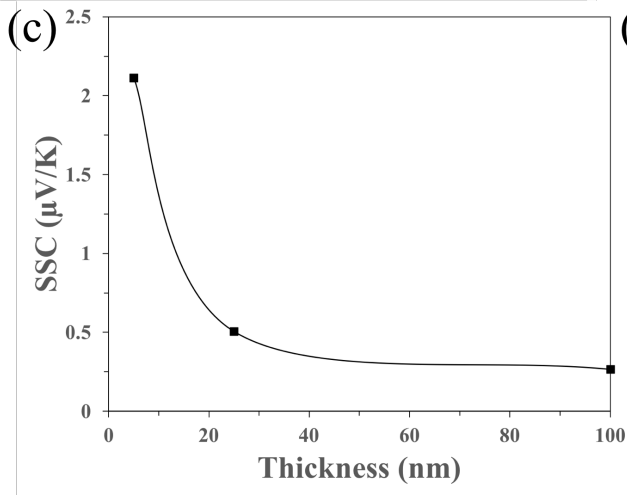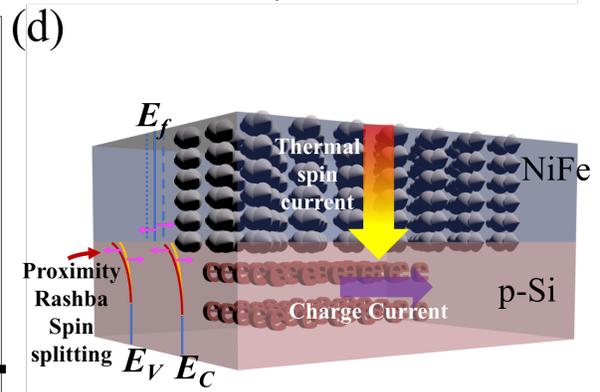